\documentclass[prb,twocolumn,showkeys,showpacs,amsmath,amsfonts,floatfix,superscriptaddress,nofootinbib]{revtex4-2}

\usepackage{amssymb,amsmath,amstext}                
\usepackage{graphicx}                                               
\usepackage{epstopdf}                                               
\usepackage{color}       
\usepackage{bm}
\usepackage{appendix}                                              
\usepackage[latin2]{inputenc}
\usepackage{ulem}
\usepackage{latexsym}
\usepackage[colorlinks=true,citecolor=blue,linkcolor=magenta]{hyperref}
\def\be{\begin{equation}}
\def\ee{\end{equation}}
\def\bea{\begin{eqnarray}}
\def\eea{\end{eqnarray}}
\def\bi{\begin{itemize}}
\def\ei{\end{itemize}}

\newcommand{\bra}[1]{\mbox{$\langle #1 |$}}
\newcommand{\ket}[1]{\mbox{$| #1 \rangle$}}
\newcommand{\braket}[2]{\mbox{$\langle #1  | #2 \rangle$}}

\begin{document}

\title{  Efficient Representation of Minimally Entangled Typical Thermal States \\ in two dimensions via Projected Entangled Pair States}

\author{Aritra Sinha} 
\affiliation{Jagiellonian University, 
             Faculty of Physics, Astronomy and Applied Computer Science,
             Institute of Theoretical Physics, 
             ul. \L{}ojasiewicza 11, 30-348 Krak\'ow, Poland}  
\affiliation{Max Planck Institute for the Physics of Complex Systems, 
             N\"{o}thnitzer Strasse 38, Dresden 01187, Germany}
             
\author{Marek M. Rams} 
\affiliation{Jagiellonian University, 
             Faculty of Physics, Astronomy and Applied Computer Science,
             Institute of Theoretical Physics, 
             ul. \L{}ojasiewicza 11, 30-348 Krak\'ow, Poland}  
\affiliation{Jagiellonian University, 
             Mark Kac Center for Complex Systems Research,
             ul. \L{}ojasiewicza 11, 30-348 Krak\'ow, Poland}  

\author{Jacek Dziarmaga} 
\affiliation{Jagiellonian University, 
             Faculty of Physics, Astronomy and Applied Computer Science,
             Institute of Theoretical Physics, 
             ul. \L{}ojasiewicza 11, 30-348 Krak\'ow, Poland}  
\affiliation{Jagiellonian University, 
             Mark Kac Center for Complex Systems Research,
             ul. \L{}ojasiewicza 11, 30-348 Krak\'ow, Poland}

\date{\today}

\begin{abstract}
The Minimally Entangled Typical Thermal States (METTS) are an ensemble of pure states, equivalent to the Gibbs thermal state, designed with an efficient tensor network representation in mind. In this article, we use the Projected Entangled Pair States (PEPS) as their representation on a two-dimensional (2D) lattice. Unlike Matrix Product States (MPS), which for 2D systems are limited by an exponential computational barrier in the lattice size, PEPS provides a more tractable approach. 
To substantiate the prowess of PEPS in modeling METTS (dubbed as PEPS-METTS), we benchmark it against the purification method in the context of the 2D quantum Ising model at its critical temperature. Our analysis reveals that PEPS-METTS achieves accurate results with significantly lower bond dimensions. We further corroborate this finding in the 2D Fermi-Hubbard model.
At a technical level, we introduce an efficient \textit{zipper} method to obtain PEPS boundary MPS needed to compute expectation values and perform sampling. The imaginary time evolution is done with the neighbourhood tensor update.
\end{abstract}

\maketitle

\section{Introduction}
\label{sec:intro}

Tensor networks have become indispensable tools in the computational study of condensed matter physics, allowing for efficient representations of quantum states~\cite{Verstraete_review_08, Orus_review_14}. They include the one-dimensional (1D) matrix product state (MPS)~\cite{fannes1992} and 2D projected entangled pair states (PEPS)~\cite{Nishino_2DvarTN_04, verstraete2004}. MPS can represent ground states of 1D local Hamiltonians~\cite{Verstraete_review_08, Hastings_GSarealaw_07, Schuch_MPSapprox_08} and their thermal states~\cite{Barthel_1DTMPSapprox_17}. It serves as the variational ansatz underlying the famous density matrix renormalization group (DMRG)~\cite{White_DMRG_92, White_DMRG_93, Schollwock_review_05, Schollwock_review_11}. 
It is also expected that a 2D PEPS forms a good variational ansatz for ground and thermal states of similar 2D Hamiltonians~\cite{Verstraete_review_08, Orus_review_14, Wolf_Tarealaw_08, Molnar_TPEPSapprox_15}, although its ability to represent 2D states satisfying the area law has its limitations~\cite{Eisert_TNapprox_16, HasikChiral}. As tensor networks do not suffer from the sign problem notorious in the quantum Monte Carlo, they can treat fermionic systems~\cite{Corboz_fMERA_10, Eisert_fMERA_09, Corboz_fMERA_09, Barthel_fTN_09, Gu_fTN_10}, as demonstrated for both finite~\cite{Cirac_fPEPS_10} and infinite PEPS (iPEPS)~\cite{Corboz_fiPEPS_10, Corboz_stripes_11}.

PEPS was initially proposed to represent ground states of finite systems~\cite{Nishino_2DvarTN_04, verstraete2004, Murg_finitePEPS_07}. With the advent of efficient algorithms it was upgraded to an infinite PEPS ~\cite{Cirac_iPEPS_08,Xiang_SU_08,Gu_TERG_08,Orus_CTM_09}, which proved to be one of the methods of choice for strongly correlated quantum systems in 2D. 
The method has been pivotal in several groundbreaking applications. For instance, it unraveled the enigmatic magnetization plateaus in the complex compound $\textrm{SrCu}_2(\textrm{BO}_3)_2$~\cite{matsuda13,corboz14_shastry}. Additionally, it provided strong evidence of the stripy nature of the ground state in the doped 2D Hubbard model~\cite{Simons_Hubb_17}. Moreover, it sheds light on the existence of gapless spin liquid phase in the kagome Heisenberg antiferromagnet~\cite{Xinag_kagome_17}. 
Subsequent advancements in the field~\cite{fu,Corboz_varopt_16, Vanderstraeten_varopt_16, Fishman_FPCTM_17, Xie_PEPScontr_17, Corboz_Eextrap_16, Corboz_FCLS_18, Rader_FCLS_18, Rams_xiD_18} have set the stage for simulating thermal states~\cite{Czarnik_evproj_12, Czarnik_fevproj_14, Czarnik_SCevproj_15, Czarnik_compass_16, Czarnik_VTNR_15, Czarnik_fVTNR_16, Czarnik_eg_17, Dai_fidelity_17, CzarnikDziarmagaCorboz, czarnik19b, Orus_SUfiniteT_18, CzarnikKH, jimenez20, Poilblanc_thermal, CzarnikSS, Poilblanc_thermal, gauthe2022thermal, ntuHubbard,gauthe2023thermal}, mixed states in open systems~\cite{Kshetrimayum_diss_17, CzarnikDziarmagaCorboz, SzymanskaPEPS}, excited states~\cite{Vanderstraeten_tangentPEPS_15, ExcitationCorboz, XiangTriangularSpectra}, and even real-time dynamics~\cite{CzarnikDziarmagaCorboz, HubigCirac, tJholeHubig, SUlocalization, SUtimecrystal, ntu, KZ2D, BH2Dcorrelationspreading, mbl_ntu, Poilblanc_evolution}.

Tensor network alternatives to iPEPS are also under constant development, e.g., simulating systems on infinite cylinders or finite lattices using MPS. Thanks to its stability, this method is now routinely used to investigate 2D ground states ~\cite{Simons_Hubb_17, CincioVidal} and was also applied to thermal states~\cite{Stoudenmire_2DMETTS_17, Weichselbaum_Tdec_18, WeichselbaumTriangular, WeichselbaumBenchmark, stripes_finite_mets, Wei2021, li2023tangent}. However, this approach is sharply limited by the exponential growth of the computational complexity with the system size, given by requirement on its refinement parameter, \textit{bond dimension}.  Alternative approaches include direct renormalization of a $3$D tensor network representing a $2$D thermal density matrix~\cite{Li_LTRG_11, Xie_HOSRG_12, Ran_ODTNS_12, Ran_NCD_13, Ran_THAFstar_18, Su_THAFoctakagome_17, Su_THAFkagome_17, Ran_Tembedding_18}, cluster expansion~\cite{cluster_thermal} or the recently developed isometric tensor network states~\cite{kadow2023isometric}.

In a typical approach, a purification of a thermal state is represented by a tensor network encompassing the whole thermal ensemble in a compact way. The representation is efficient at high temperatures, where the purification is weakly entangled. However, at sufficiently low temperature it becomes isomorphic to a tensor square of the ground state. In other words, the complexity of the network, as measured by its bond dimension, is squared in comparison to what is needed for just the ground state when using the same PEPS scheme. This exemplifies the problem of potentially sub-optimal representation, given that at zero temperature the entire thermal ensemble could be described by just the ground state itself.

Prompted by the computational complexities and limitations of existing tensor network methods, we propose an alternate methodology: the fusion of finite PEPS with Minimally Entangled Typical Thermal States (METTS)~\cite{METTS_White, Stoudenmire_2DMETTS_17}. The aim is to synergistically combine METTS' computational efficiency with the representational capacity of PEPS. For sufficiently low temperatures, METTS closely approximates the ground state. Thereby, it should provide a computationally tractable means to explore intricate phenomena, such as stripe formations in the Hubbard model, observed, for instance, in MPS simulations on narrow cylinders~\cite{stripes_finite_mets, METTS_triangular}. Our approach hopes to take this a step further by incorporating the true 2D ansatz PEPS, effectively broadening the applicability of METTS to 2D systems. It has the potential to be a methodological leap that could bypass the existing challenges tied to the use of MPS for intrinsically 2D systems.

The rest of the paper is organized as follows. In Sec.~\ref{sec:metts}, we outline the METTS stochastic unraveling of the thermal state~\cite{METTS_White, METTS_NJP, Stoudenmire_2DMETTS_17, stripes_finite_mets, METTS_triangular}. The METTS algorithm iterates imaginary time evolution followed by a projective measurement. The evolution step, calculation of expectation values, and the projective measurements are described in Sec.~\ref{sec:peps}. The latter two require the PEPS's norm boundary, efficiently calculated with a zipping procedure in Sec.~\ref{sec:zip}. We benchmark PEPS-METTS against the standard purification method, outlined in Sec.~\ref{sec:purification}. Benchmark results for the 2D quantum Ising model are presented in Sec.~\ref{sec:QIM} and those for the Hubbard model in Sec.~\ref{sec:Hubbard}. We conclude in Sec.~\ref{sec:conclusion}. Our primary contributions in this paper include integration of METTS with PEPS, benchmarking the results against the standard purification method for two pivotal models, and a future outlook that suggests promising avenues for methodological and computational advances.

\section{METTS}
\label{sec:metts}

\begin{figure}[t!]
\includegraphics[width=0.95\columnwidth,clip=true]{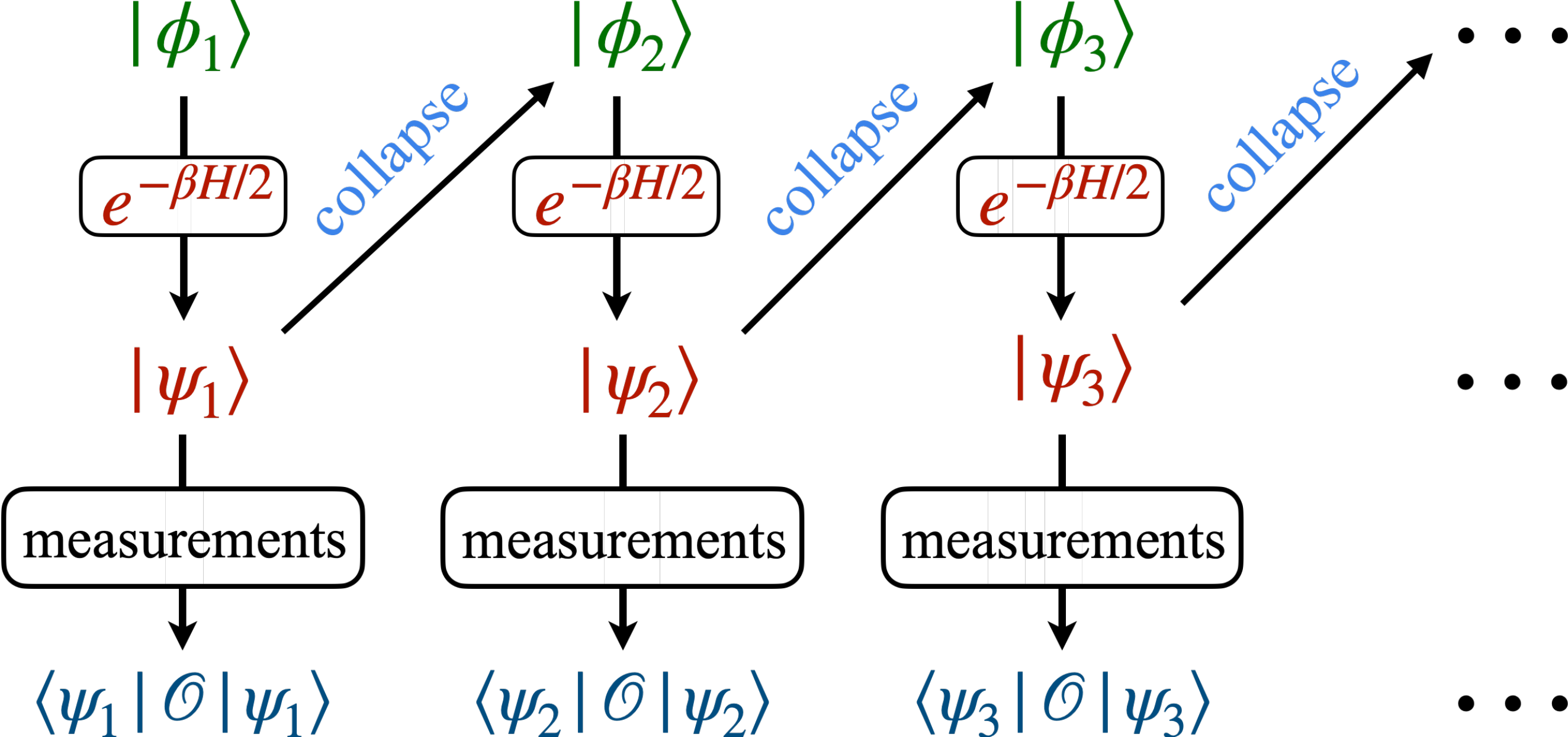}
\caption{
{\bf Minimally entangled typical thermal states.}
The flowchart depicts the METTS algorithm, detailed in Sec.~\ref{sec:metts}. A product state $|\phi_{i}\rangle$ undergoes imaginary-time evolution for time $\beta/2$, transforming it into METTS states $|\psi_{i}\rangle$. Its contribution to the thermal expectation value of interest is given by $\langle \psi_{i}| \mathcal{O}|\psi_{i}\rangle$. Subsequently, the next product state is randomly sampled from $|\psi_{i}\rangle$, and the procedure starts anew, resulting in a Markov chain. Finally, the expectation value of interest is estimated as an average over the obtained measurement results. In this work, the states are represented using PEPS ansatz, with the evolution performed with the NTU algorithm.
}
\label{fig:cartoonMETTS}
\end{figure}

The METTS algorithm can be summarized as follows~\cite{METTS_White, METTS_NJP, Stoudenmire_2DMETTS_17, stripes_finite_mets, METTS_triangular}. With an orthonormal basis, $\{\ket{\phi_i}\}$, the thermal average of an operator ${\cal O}$ can be written as 
\bea 
\langle \cal{O} \rangle 
&=&
\frac{1}{\cal{Z}}
\sum_i
\bra{\phi_i} e^{-\beta H/2} {\cal O} e^{-\beta H/2} \ket{\phi_i} \nonumber\\ 
&=&
\sum_i p_i \bra{\psi_i} {\cal O} \ket{\psi_i}.
\eea
Here $p_i=\bra{\phi_i}e^{-\beta H}\ket{\phi_i}$, and $\ket{\psi_i}=p_i^{-1/2} e^{-\beta H/2}\ket{\phi_i}$ are normalized typical thermal states. After normalization, their weights $p_i$ become a probability distribution: ${\cal Z}^{-1}\sum_i p_i=1$, where ${\cal Z}$ is the statistical sum.
Assuming ergodicity, the thermal average follows from the Monte Carlo sampling over a Markov chain with a stationary distribution $p_i$. The thermal average is estimated as an average
\be 
\langle {\cal O} \rangle = \lim_{s\to\infty} \frac{1}{s} \sum_{j=1}^s {\cal O}_j.
\label{eq:running_avarage}
\ee 
Here, 
\be 
{\cal O}_j=\bra{\psi_j} {\cal O} \ket{\psi_j}
\label{eq:calOi}
\ee
is the expectation value of the operator in the $j$-th typical thermal state. The numerical algorithm repeats a sequence of the imaginary time evolution, $e^{-\beta H/2}$, followed by calculation of the expectation value, ${\cal O}_j$, and then projection on the orthonormal basis $\{\ket{\phi_i}\}$, as summarized in Fig.~\ref{fig:cartoonMETTS}.

A product basis is a convenient choice, not only to make the projective measurement but also to perform the subsequent imaginary time evolution of the state represented by a tensor network. The initial collapsed state is a product over the lattice sites and can be represented by a trivial tensor network with bond dimension one. The evolution that follows builds correlations and increases the bond dimension. The latter can in principle be further minimized by optimizing the product basis, though the choice of the basis might also affect the ergodicity. All this makes tensor networks a natural representation for the minimally entangled states, but up till now, only MPS were employed, either in 1D~\cite{METTS_White, METTS_NJP} or on thin cylinders~\cite{Stoudenmire_2DMETTS_17, stripes_finite_mets, METTS_triangular}. Thickening the cylinder towards a truly 2D lattice is limited by the exponential growth of the MPS bond dimension. It motivates our attempt in the 2D setups to employ PEPS instead, which is a genuine 2D tensor network.

\begin{figure}[t!]
\includegraphics[width=\columnwidth,clip=true]{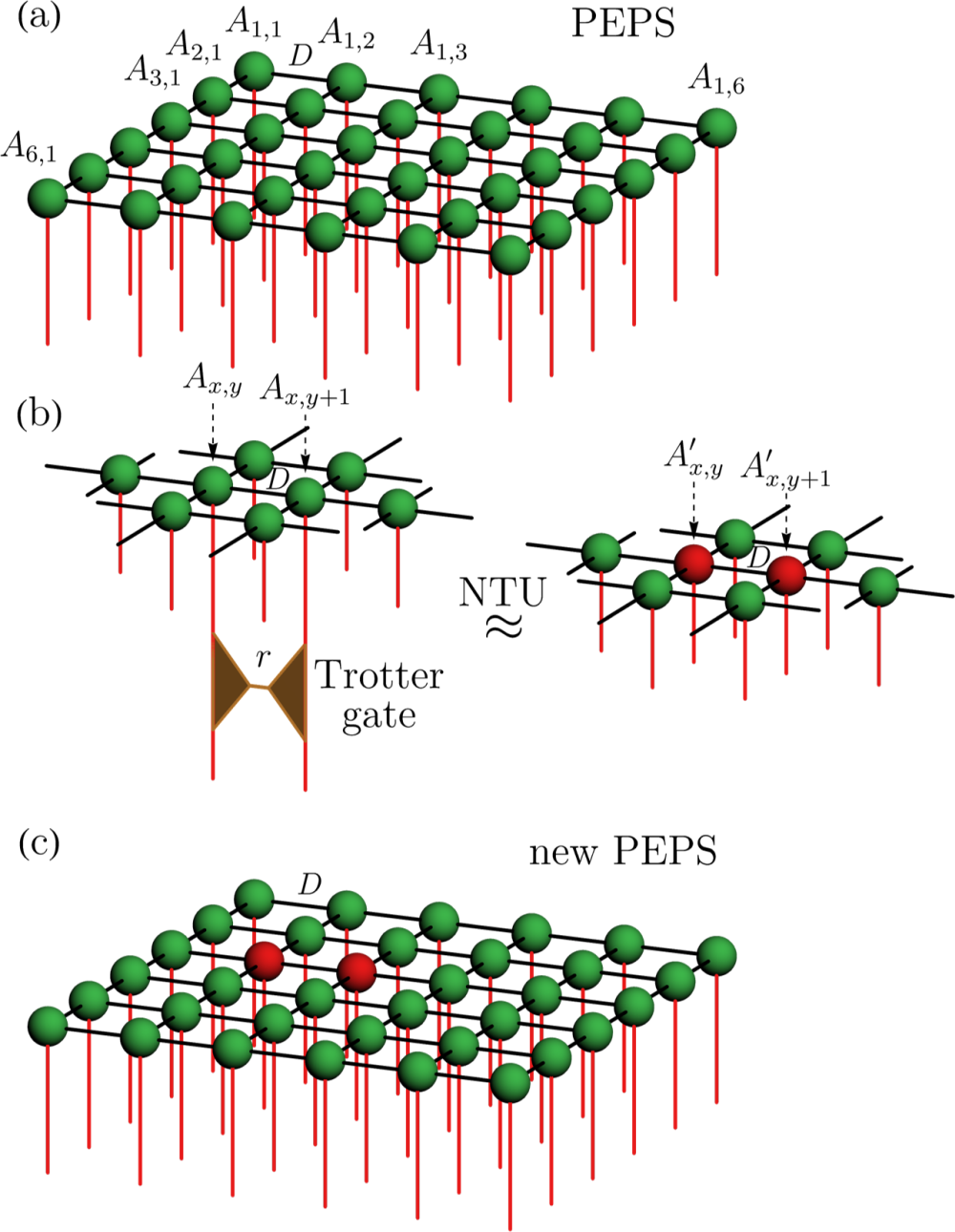}
\caption{
{\bf Evolution by NTU.}
In (a), finite PEPS representing a quantum state on a $6\times 6$ open-boundary square lattice is a contraction of PEPS tensors $A_{x,y}$. Each bond index (black lines), connecting a pair of nearest neighbor (NN) tensors, has the same bond dimension $D$. The red lines represent physical indices.
In each Suzuki-Trotter step, a Trotter gate is applied to a NN pair of PEPS tensors. The gate can be represented by a contraction of two tensors by an index with dimension~$r$. When the two tensors are absorbed into the original PEPS tensors, the bond dimension between them increases from $D$ to $r\times D$. It has to be truncated back to the original $D$.
For instance, in (b), a horizontal pair of NN PEPS tensors: $A_{x,y}$ and $A_{x,y+1}$ -- with a Trotter gate applied to it -- is approximated by a pair of new (red) PEPS tensors $A'_{x,y}$ and $A'_{x,y+1}$ connected by an index with the original dimension $D$. The new tensors are optimized to minimize the Frobenius norm of the difference between the two networks in panel (b).
In (c), the optimized new PEPS tensors replace the original pair of tensors in the new PEPS, here for $(x,y)=(3,2)$. 
Now, the next Trotter gate can be applied.
Instead of the sequential application, NN gates can also be applied in parallel on non-overlapping NTU clusters. 
}
\label{fig:NTU_overview}
\end{figure}

\begin{figure}[t!]
\includegraphics[width=0.95\columnwidth,clip=true]{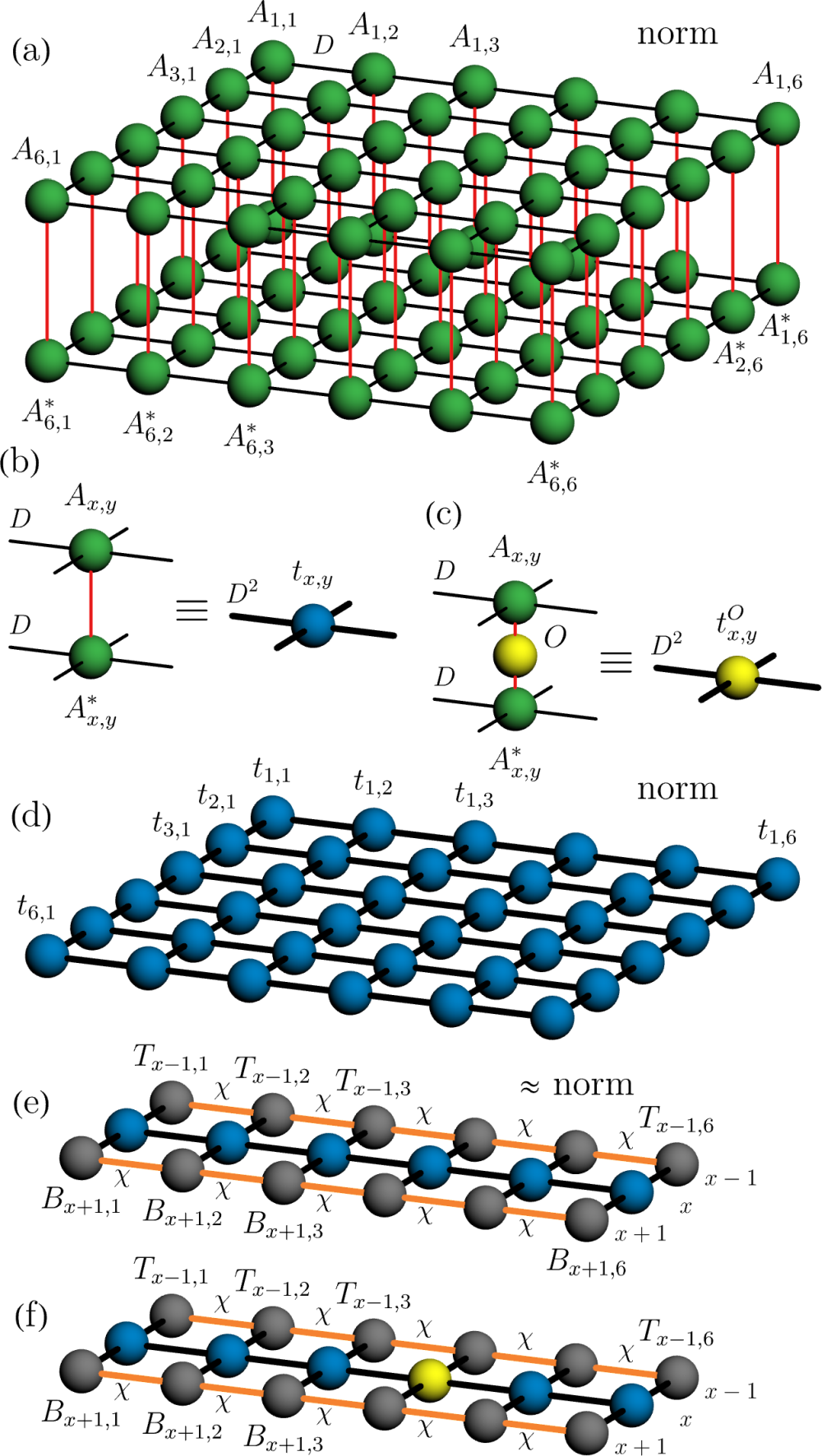}
\caption{
{\bf Expectation value.}
Panel (a) shows contraction of the PEPS~$\ket{\psi}$ in Fig.~\ref{fig:NTU_overview}(a) (top layer) with its conjugate $\bra{\psi}$ (bottom layer) into a squared norm $\braket{\psi}{\psi}$. Each PEPS tensor $A_{x,y}$ can be contracted with its conjugate $A^*_{x,y}$ into a transfer tensor, $t_{x,y}$, as shown in (b). With operator $O$ at site $(x,y)$, the tensor becomes $t^O_{x,y}$, as shown in (c). In terms of the transfer tensors, the norm in (a) becomes the planar diagram in (d). In (e), the top rows $1,\ldots,x-1$ in the exact diagram (d) are approximated by a boundary matrix product state (MPS) with tensors $T_{x-1,y}$, here in right--canonical form. Similarly, the bottom rows $x+1,\ldots,L$ are approximated by a boundary MPS with tensors $B_{x+1,y}$. Both boundary MPSs have bond dimension $\chi$. In (f), the expectation value of operator $O$ at site $(x,4)$ is calculated. For an unnormalized PEPS, it has to be normalized by (e).  
}
\label{fig:exp}
\end{figure}

\section{PEPS evolution by NTU}
\label{sec:peps}

There are well-established iPEPS techniques for translationally invariant states on an infinite lattice~\cite{Cirac_iPEPS_08, Xiang_SU_08, Gu_TERG_08, Orus_CTM_09}. In the METTS context, however, projective measurements break lattice symmetries, making it more natural to work with a PEPS ansatz on a finite lattice. The network is shown in Fig.~\ref{fig:NTU_overview}(a). 

Its time evolution is performed by the second-order Suzuki-Trotter decomposition into small time steps. An application of the nearest neighbor (NN) two-site Trotter gate is outlined in Figs.~\ref{fig:NTU_overview}(b) and (c). The gate increases the bond dimension on the NN bond. In order to prevent exponential growth of the bond dimension during the time evolution, the increased bond dimension has to be truncated back to a predefined maximal value $D$.
In this work, the truncation is done as in Fig.~\ref{fig:NTU_overview}(b), where a cluster including the NN bond with a gate and its neighboring sites is approximated by a similar cluster but with dimension $D$ of the bond. Minimization of the Frobenius norm of the difference between the two diagrams is the essence of the neighborhood tensor update (NTU) algorithm~\cite{ntu}. NTU can be regarded as a special case of a cluster update~\cite{wang2011cluster}, where the cluster size is a refinement parameter interpolating between a local and an infinite cluster.

In the case of ground-state calculations, an interplay between the maximal achievable correlation length and the cluster size was demonstrated~\cite{Lubasch_cluster_1, Lubasch_cluster_2}. However, in the case of NTU, the small cluster is chosen such that the Frobenius norm is calculated numerically exactly. It yields a manifestly non-negative and Hermitian effective metric tensor used for truncation of PEPS tensors and warrants the algorithm's stability. The usefulness of NTU was already demonstrated in the Kibble-Zurek quenches in 2D~\cite{KZ2D, KZBH2D} or unitary time evolution of many-body localizing systems after a sudden quench~\cite{mbl_ntu}. In the context of thermal states, NTU was benchmarked by the imaginary time evolution of a thermal purification of the 2D quantum Ising model represented by iPEPS~\cite{ntu}. The same technique was used to address the fermionic Hubbard model on an infinite square lattice at medium and high temperatures~\cite{ntuHubbard}. It seems a reasonable choice for weakly entangled typical thermal states.

\begin{figure}[t!]
\includegraphics[width=0.999\columnwidth,clip=true]{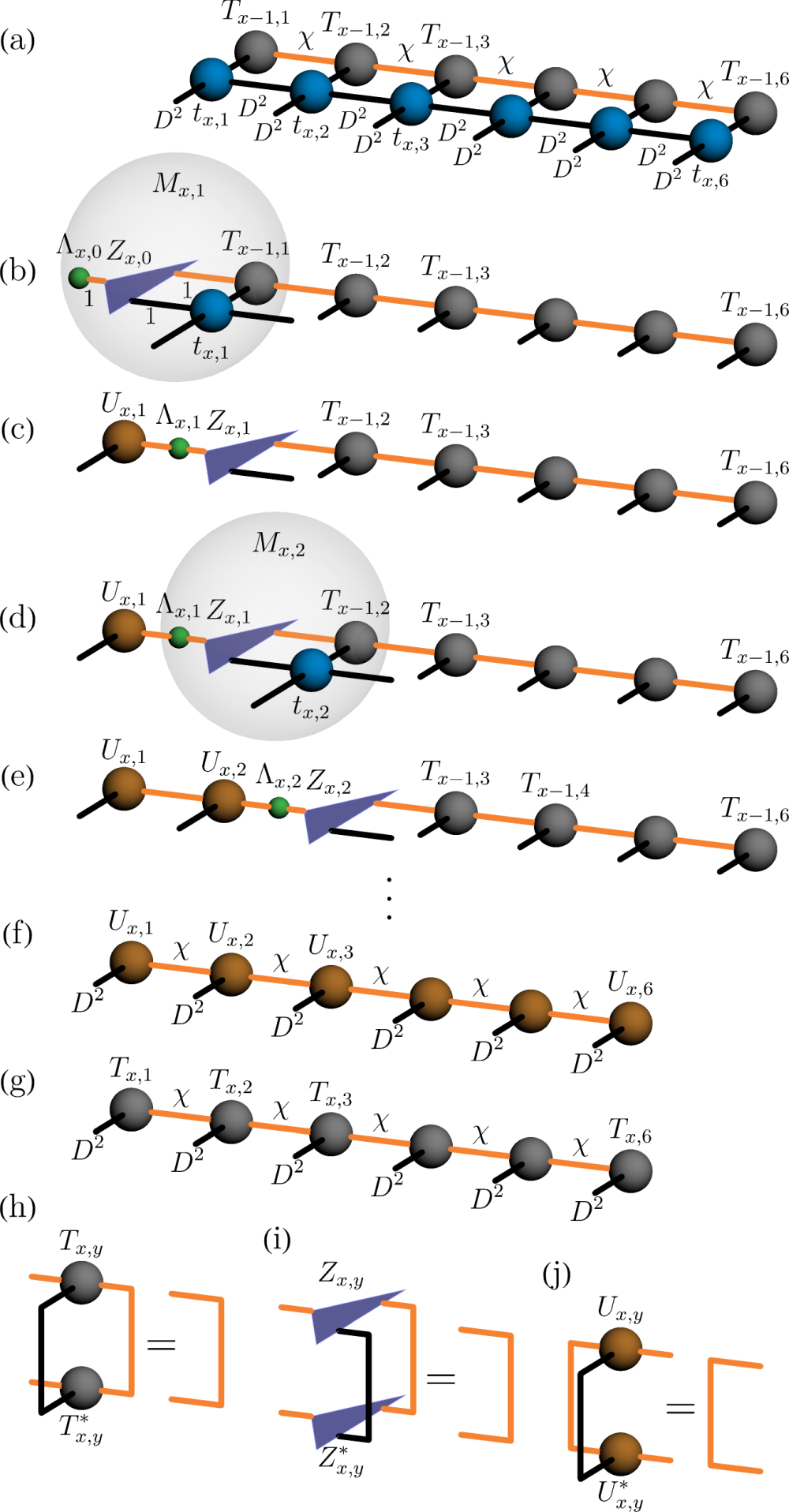}
\caption{
{\bf Zipping PEPS boundary.}
This figure shows step by step how to approximate the diagram in~(a)---the $(x-1)$-st PEPS boundary MPS made of tensors $T_{x,y}$ multiplied by the $x$-th row transfer matrix made of tensors $t_{x,y}$---with the $x$-th boundary MPS with bond dimension $\chi$ in (g). The transfer matrix is not applied at once. Instead, we sweep through the MPS applying one tensor $t_{x,y}$ at a time accompanied by SVD truncation of the bond dimension, see panels (b--e). This way, the $(x-1)$-st MPS is zipped with the row transfer matrix into the left--canonical MPS in (f), which can be brought back to the right--canonical form in (g).
}
\label{fig:zip}
\end{figure}

\begin{figure}[t!]
\includegraphics[width=0.999\columnwidth,clip=true]{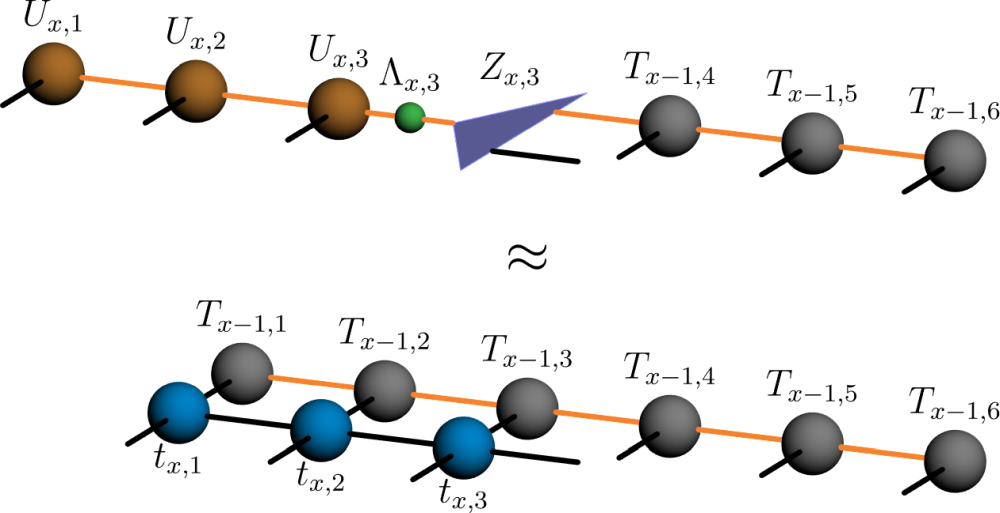}
\caption{
{\bf Mixed PEPS boundary.}
Two approximately equal representations of the mixed PEPS boundary which arise midway through the zipping.  The mixed boundary connects a segment of the next $x$-th boundary with a segment of the previous $(x-1)$-st boundary. It has an extra horizontal ``physical'' bond of dimension $D^2$ that disappears after the zipper is swept all the way to the right edge. 
}
\label{fig:zip_bis}
\end{figure}

The imaginary time evolution is followed by calculation of expectation values ${\cal O}_j$ in Eq.~\eqref{eq:calOi}. Fig.~\ref{fig:exp} depicts calculation of an expectation value for a one-site operator ${\cal O}$. Towards this end, the PEPS norm in Fig.~\ref{fig:exp}(a) is replaced by a planar network in (d). The top rows above the site are approximated by a top boundary MPS and the bottom rows by a bottom boundary MPS. As a result, one obtains quasi-1D diagrams in panels (e) and (f), which can be contracted numerically exactly. There are many techniques to obtain the boundary MPSs, see e.g.~\cite{tangent_review}, but in this work, we introduce an efficient zipper method described in the next section and illustrated by diagrams in Figs.~\ref{fig:zip} and \ref{fig:zip_bis}. One of the alternatives is to apply the corner transfer matrix renormalization group~\cite{Orus_CTM_09}.

\begin{figure}[t!]
\includegraphics[width=\columnwidth,clip=true]{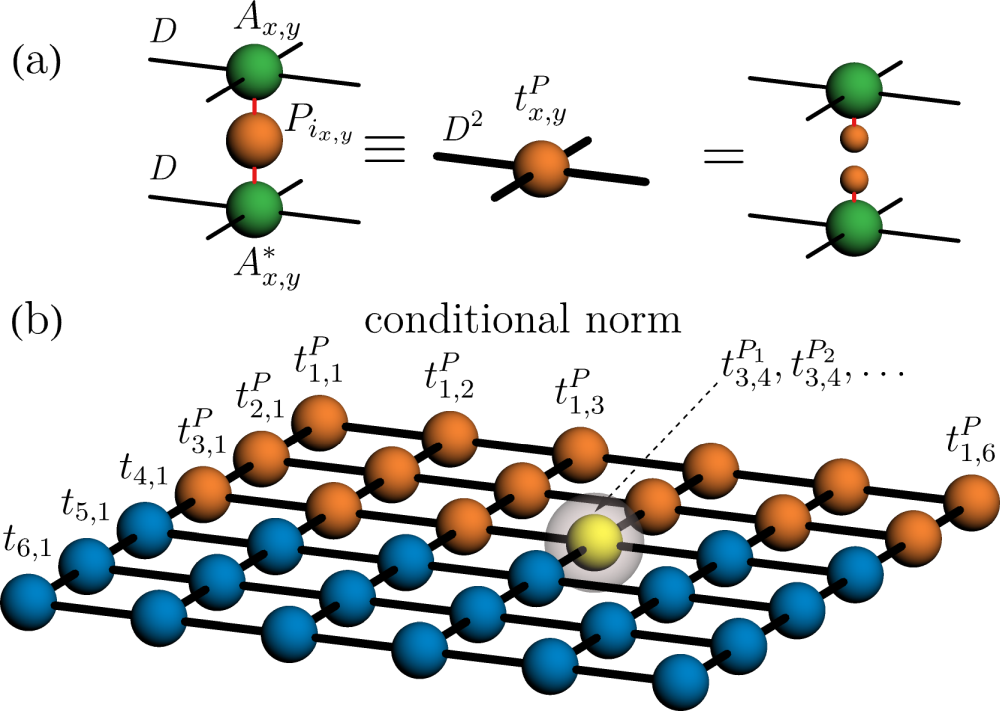}
\caption{
{\bf Sampling.}
In panel (a), a transfer tensor at site $(x,y)$ is inserted with a projector $P_{i_{x,y}}$ on the measurement outcome. The projector can be split into two vectors to separate the top and bottom layers of PEPS tensors.
A single layer can be used to contract the upper part of the network in (b) more efficiently.
Panel (b) shows the conditional probability for possible measurement outcomes at site $(x,y)=(3,4)$. The first two rows and the first three sites of the third row were already measured, and the PEPS was updated with the projectors corresponding to the actual measurement outcomes. At site $(3,4)$, all orthogonal projectors $P_i=\ket{\sigma_i}\bra{\sigma_i}$ get probed. An expectation value of each projector, equal to the corresponding outcome probability, is calculated as in Fig.~\ref{fig:exp}. It allows drawing the measurement outcome on that site according to obtained probabilities. Repeating this procedure site-after-site, going row after row, allows sampling from the PEPS state in the computational product basis.
}
\label{fig:collapse}
\end{figure}

Once all expectation values of interest are calculated, we can proceed to projective measurement in the product basis $\{\ket{\phi_i}\} = \{\ket{\sigma_{i_{1,1}} \sigma_{i_{1,2}} \sigma_{i_{1,3}} \ldots}\}$.  It is convenient to perform the measurement sequentially site by site, similar to sampling from a classical PEPS representing thermal state~\cite{ueda2005snapshot,rams2021approximate,FriasPerez2023collective}. Fig.~\ref{fig:collapse} shows how to calculate outcome probabilities when the sites are measured row by row from left to right. The first measurement is done at site $(1,1)$. Outcome probabilities are given by the expectation values of all orthogonal projectors at this site, $P_{i_{1,1}}$, calculated as in Fig.~\ref{fig:exp}. The measurement outcome is selected randomly according to the obtained probabilities, and the PEPS state gets updated with the projector $P_{i_{1,1}}$ corresponding to the actual measurement outcome. 

Fig.~\ref{fig:collapse}(b) shows the probability distribution for different measurement outcomes at site $(3,4)$, conditioned on the measurement outcomes in the first two rows and the first three sites in the third row. As compared to the norm before these measurements, see Fig.~\ref{fig:exp}(d), the transfer tensors on the measured sites were replaced by new transfer tensors inserted with the measurement projectors, see Fig.~\ref{fig:collapse}(a), and the transfer tensors at site $(3,4)$ with projectors corresponding to all possible measurement outcomes at this site. Expectation values of the latter projectors, calculated in the same way as the expectation value in Figs.~\ref{fig:exp}(d)--(f), give the conditional outcome probabilities at site $(3,4)$. Note that projectors in the top PEPS transfer matrices in Fig.~\ref{fig:collapse}(b) make the top and bottom PEPS tensors in Fig.~\ref{fig:collapse}(a) disjoined. As such, for those rows, it is sufficient to work with a single-layer of PEPS tensors approximating its boundary MPS.

Finally, after all the sites have been measured, a new PEPS with bond dimension one is initialized in a product state $\ket{\sigma_{i_{1,1}} \sigma_{i_{1,2}} \sigma_{i_{1,3}}\ldots}$ corresponding to the drawn measurement outcome---and the imaginary time evolution starts again. In this sense, every measurement resets the bond dimension back to one, keeping the states minimally entangled. 

\section{ Zipping PEPS boundary.}
\label{sec:zip}

Fig.~\ref{fig:zip} outlines the main steps to obtain the top PEPS boundary that we need in Figs.~\ref{fig:exp}(e) and (f). The bottom boundary is obtained similarly. The top tensors are right--canonical isometries, see Fig.~\ref{fig:zip}(h). Here, the overall task is to approximate the diagram in Fig.~\ref{fig:zip}(a)---where the $x$-th row of transfer tensors is applied to the $(x-1)$-st boundary---with the $x$-th boundary in Fig.~\ref{fig:zip}(f).

Breaking away from the usual approach, we abstain from applying the whole row of $t$'s in Fig.~\ref{fig:zip}(a) at once. Instead, we attach them one-by-one starting from the leftmost $t_{x,1}$, see Fig.~\ref{fig:zip}(b). 
In each step of the sweep, the tensors $\Lambda_{x,y-1}$, $Z_{x,y-1}$, $T_{x-1,y}$ and $t_{x,y}$ are contracted and reshaped into matrix $M_{x,y}$ of dimensions $\chi D^2$, see panels (b) and (d) for $y=1$ and $2$, respectively ($\Lambda_{x,0}$, $Z_{x,0}$ are initialized as unit tensors of dimensions one). In panels (c) and (e), matrix $M_{x,y}$ is replaced by its singular value decomposition, $M_{x,y}=U_{x,y}\Lambda_{x,y}Z_{x,y}$, where, by construction, $U$ is a left-canonical isometry, see panel (j), $Z$ is a right--canonical one, see panel (i), and $\Lambda$ is a diagonal matrix of singular values.
At this point, we make an approximation by truncating the total of $\chi D^2$ singular values down to the $\chi$ leading ones. 
The complete sweep results in the left--canonical MPS with bond dimension $\chi$ in panel (f), which approximates the MPO-MPS product in panel (a). Note, that during each SVD truncation, all MPS tensors to the left(right) of $M_{x,y}$ are in left(right)--canonical form, making the truncation locally optimal.

The final cosmetic move brings the final MPS back to the right--canonical form in panel (g). This completes the application of the transfer matrix made of the $x$-th row of transfer tensors. Now the next $(x+1)$-st row can be applied to obtain the $(x+1)$-st top boundary in the same way (alternatively, one can start in the left--canonical form in panel (f) and zip in the opposite direction).
As a brief summary, Fig.~\ref{fig:zip_bis} shows two approximately equivalent forms of a mixed boundary midway through the zipping. 

The numerical cost of the procedure is dominated by performing SVD, with $D^6 \chi^3$ complexity (this can be reduced by employing truncated SVD, however,  typically at a cost of a significant loss of precision). For comparison, the numerical cost of SVD truncation of the entire MPS in panel (a) would scale as $D^8 \chi^3$.

The accuracy of the $x$-th boundary in Fig.~\ref{fig:zip}(f) [or (g)] can be further improved, treating it as an initial guess for the standard variational scheme~\cite{Verstraete_review_08} that maximizes the overlap of truncated MPS in panel (g) with the untruncated MPO-MPS product in panel (a). In the present work, however, it was not necessary to resort to that.


\section{ Purification}
\label{sec:purification}

In this work, we benchmark METTS against the more standard purification approach, where each physical spin is accompanied by an ancilla partner \cite{Czarnik_evproj_12}. An initial purification at $\beta=0$ is a product state over lattice sites:
\be 
\ket{\psi(0)}=
\prod_k
\left(
\sum_{i_k=1}^d
\ket{\sigma_{i_k},\sigma_{i_k}}
\right).
\label{eq:psi0}
\ee
Here $d$ is the dimension of the local physical Hilbert space, the product is over all lattice sites, and the state $\ket{\sigma_{s_{k}},\sigma_{a_{k}}}$ denotes the product of the $s_{k}$ physical basis state and $a_{k}$ ancilla basis state at site $k$.

\begin{figure}[t]
\includegraphics[width=0.9\columnwidth,clip=true]{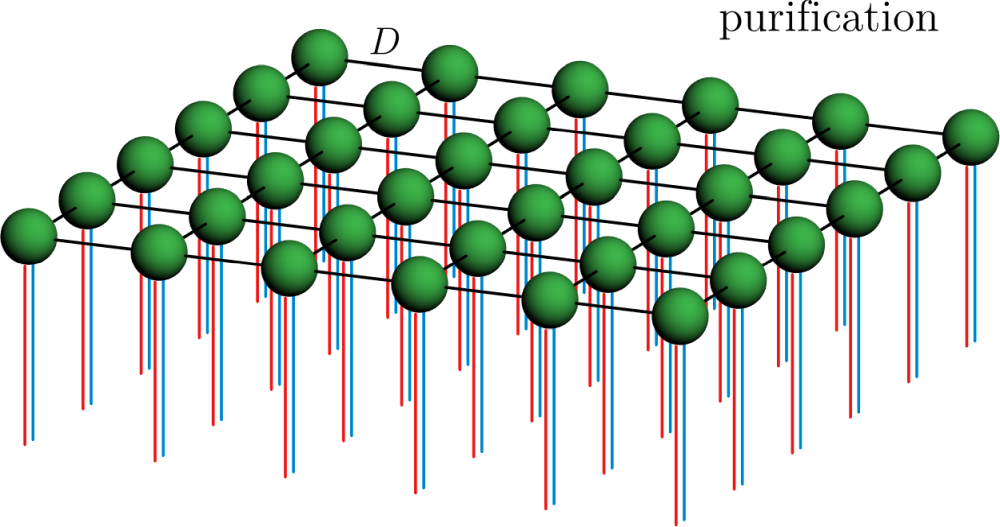}
\caption{
{\bf Purification.}
A finite PEPS representing a purification of a thermal state on a $6\times 6$ open--boundary square lattice. Each bond index (black line) connecting a pair of NN tensors has the same bond dimension $D$. The red(blue) lines represent physical(ancilla) indices.
}
\label{fig:purification}
\end{figure}

The initial state evolves into
\be 
\ket{\psi(\beta)}=e^{-\beta H/2}\ket{\psi(0)},
\ee
where the Hamiltonian is acting on the physical spins only. The thermal density matrix is obtained by tracing out the ancillas,
\be 
\rho(\beta) \propto {\rm Tr}_a \ket{\psi(\beta)}\bra{\psi(\beta)}.
\ee
The purification can be represented by a PEPS in Fig.~\ref{fig:purification}. Just as for METTS, in order to put the two methods on equal footing, the evolution is performed with the NTU. The results serve as a benchmark for METTS.

The purification is isomorphic to $e^{-\beta H/2}$ and, in a gapped system at low temperature, it becomes isomorfic to a square of its ground state: $\ket{\psi_0}\ket{\psi_0}$. When the ground state has bond dimension $D$, the purification has $D^2$, while a low-temperature METTS is just the ground state with bond dimension $D$.

\begin{figure*}[t!]
\vspace{-0cm}
\includegraphics[width=2\columnwidth,clip=true]{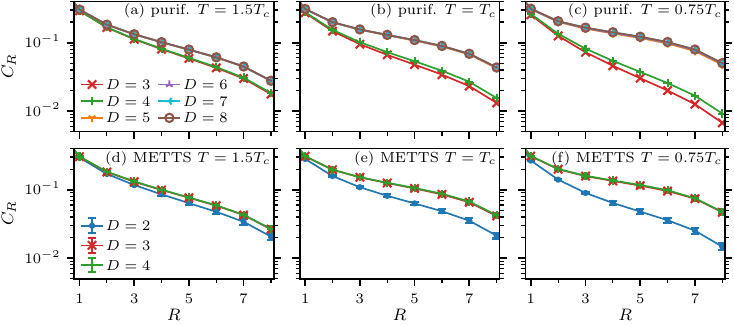}
\vspace{-0cm}
\caption{
{\bf Correlations from purification (top) and METTS (bottom) in 2D quantum Ising model.}
The ferromagnetic correlation function, $C_R=\langle \sigma^z_c \sigma^z_{c+R} \rangle$, between the central site on a $17\times 17$ open-boundary square lattice and a site at a distance $R$ along a row. Here, the transverse field $g=2.9$ and the critical temperature corresponding to this field is $T_c=0.6085$. From left to right, the temperatures are $T/T_c=1.5, 1, 0.75$.
In case of purification, panels (a), (b), and (c), for all three temperatures, the bond dimensions $D=3,4$ are too small for convergence that is reached for $D\geq5$. %
In panels (d), (e), and (f), for METTS, the error bars represent $95\%$ confidence intervals.
For all three temperatures, bond dimensions $D=3,4$ are enough to converge to the benchmark provided by the purification. The number of samples required for the error bars is $s\geq3000$.
}
\label{fig:QIM_corr_pur_metts}
\end{figure*}

\begin{figure*}[t!]
\vspace{-0cm}
\includegraphics[width=2\columnwidth,clip=true]{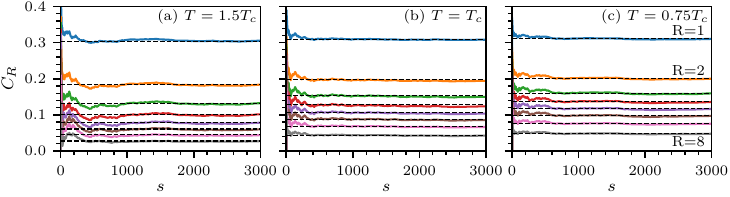}
\vspace{-0cm}
\caption{
{\bf Running average correlations.}
The ferromagnetic correlation function, $C_R=\langle \sigma^z_c \sigma^z_{c+R} \rangle$, between the central site on a $17\times 17$ open-boundary square lattice and a site at a distance $R$ along a row. 
Here, we show $\left\langle C_R \right\rangle$ in function of the length of the Markov chain, $s$. From top to bottom, $R=1,2,\ldots,8$.
The horizontal dashed lines are benchmark values from the purification.
The transverse field $g=2.9$ and the critical temperature corresponding to this field is $T_c=0.6085$. From left to right, the temperatures are $T/T_c=1.5, 1, 0.75$. 
The bond dimension is $D=3$ for METTS and $D=5$ for purification. 
}
\label{fig:QIM_corr_L17_sampling}
\end{figure*}
\begin{figure}[t!]
\vspace{-0cm}
\includegraphics[width=\columnwidth,clip=true]{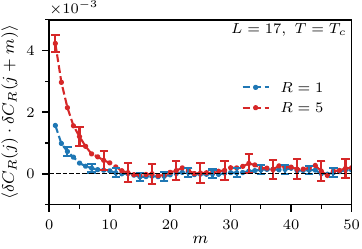}
\vspace{-0cm}
\caption{
{\bf Auto--correlations.}
We define fluctuation of the ferromagnetic correlation function as 
$\delta C_R(j)=C_R(j)-\langle C_R \rangle$, 
where $C_R(j)$ is the correlator in the $j$-th typical thermal state.
The plots show auto--correlators of the fluctuations with the error bars representing $95\%$ confidence intervals.
The auto--correlation times at $T=T_c$ are estimated~\cite{stripes_finite_mets} as $5.5$ and $7.1$ for $R=1$ and $R=5$, respectively.
Here, the bond dimension $D=4$. 
}
\label{fig:autocorr}
\end{figure}

\section{ 2D Quantum Ising Model}
\label{sec:QIM}

We test METTS by PEPS in the transverse field quantum Ising model on a square lattice with open boundary conditions,
\be 
H =
-\sum_{\langle i,j\rangle} \sigma^z_i \sigma^z_{j} - g \sum_j\sigma^x_j.
\label{eq:HIsing}
\ee
At zero temperature, the model has a ferromagnetic phase with non-zero spontaneous magnetization $\langle \sigma^z \rangle$ for the magnitude of the transverse field, $|g|$, below the quantum critical point at $g_c=3.04438(2)$~\cite{Deng_QIshc_02}. At $g=0$, the model becomes the 2D classical Ising model with a phase transition at $T_c=2/\ln(1+\sqrt{2})\approx 2.27$. In the following, we assume $g=2.9$ corresponding to a critical temperature $T_c=0.6085(8)$~\cite{Deng_QIshc_02}. When compared to $g=0$, this critical temperature is reduced almost four times by strong quantum fluctuations introduced by $g$ close to the quantum critical point. 

We test METTS by studying ferromagnetic correlation between the central site, $c$, of a $L\times L$ lattice and another site, $c+R$, at distance $R$ along its row or column. 
The correlator $C_R=\langle \sigma^z_c \sigma^z_{c+R} \rangle$ is an average over typical thermal states $\ket{\psi_j}$. The expectation value $\bra{\psi_j} \sigma^z_a \sigma^z_b \ket{\psi_j}$ is obtained along the lines of Fig.~\ref{fig:exp}.

In order to have a benchmark, first, in Fig.~\ref{fig:QIM_corr_pur_metts} (a)--(c), we obtain the correlator with the purification method for three values of temperature close to or at $T_c$. In all three cases, this approach requires $D\geq5$ for convergence. 
Next, in Fig.~\ref{fig:QIM_corr_pur_metts} (d)--(f), we obtain the same correlators with METTS. The projective measurements are done in the $\sigma^z$ basis. 
For METTS, a mere $D\geq3$ is enough to converge the correlator in the bond dimension.

This reduction of the bond dimension comes at the price of performing sampling over typical thermal states. In Fig.~\ref{fig:QIM_corr_L17_sampling}, we show estimates of the correlators in the function of the number of sampled typical thermal states,~$s$, 
\be 
C_R = 
\frac{1}{s} 
\sum_{j=1}^s 
\bra{\psi_j} \sigma^z_c \sigma^z_{c+R} \ket{\psi_j}.
\ee 
Here, for $j=1$, METTS is initialized in a state with all $\sigma^z=+1$. Somewhat counter-intuitively, at $1.5T_c$ the estimators converge with $s$ more slowly than at $T_c$, but this is just a manifestation of worse ergodicity at higher temperatures where the evolution operator, $e^{-\beta H/2}$, is slower to reshuffle consecutive measurement outcomes.

The running averages in Fig.~\ref{fig:QIM_corr_L17_sampling} may appear to have long auto--correlation tails in time, as expected even when an averaged random variable is not correlated in time at all. Indeed, we show the auto-correlators of $C_1$ and $C_5$ in Fig.~\ref{fig:autocorr}, and their auto-correlation times are around $5$--$7$ iterations. They are short, demonstrating good ergodicity. 

Although short, the finite auto--correlation times mean that the consecutive measurement outcomes are not statistically independent, and the textbook estimator of their variance would underestimate the error bars. In order to estimate statistical errors, we first bunch the data to reduce the effects of auto-correlation and then compute a running standard deviation over the running average of the bunched data.
  
\begin{figure}[t!]
\vspace{-0cm}
\includegraphics[width=0.999\columnwidth,clip=true]{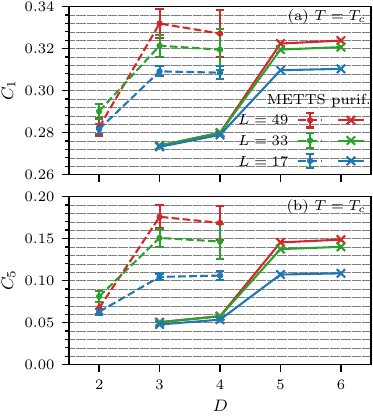}
\vspace{-0cm}
\caption{
{\bf Convergence with bond dimension.}
Top panel shows the NN ferromagnetic correlator $C_{1}$, and the bottom panel the $C_{5}$ correlator, as a function of the PEPS bond dimension for lattice sizes $L=17,33,49$.
Here, $T=T_c$ and the error bars represent $95\%$ confidence intervals.
}
\label{fig:convergence}
\end{figure}

\begin{figure}[t!]
\vspace{-0cm}
\includegraphics[width=\columnwidth,clip=true]{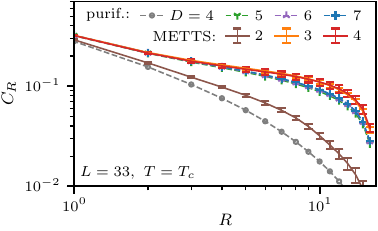}
\vspace{-0cm}
\caption{
{\bf Correlations for $\mathbf{L=33}$.}
The ferromagnetic correlation function, $C_R=\langle \sigma^z_c \sigma^z_{c+R} \rangle$, between the central site on a $33\times 33$ open-boundary square lattice and a site at a distance $R$ along the row. 
Here, $T=T_c$ and the error bars represent $95\%$ confidence intervals.
Here $D=3,4$ is enough to converge to the purification benchmark ($D=6$) only for short enough $R$, with the range increasing with $D$.
}
\label{fig:QIM_corr_L33_49_metts}
\end{figure}

Finally, in Fig. \ref{fig:convergence}, we focus on the $C_1$ and $C_{5}$ correlators at the critical temperature $T=T_c$ and show how they depends on the bond dimension $D$ for lattice sizes $L=17,33,49$. 
The NN correlator $C_1$ is well converged in $D$, both for the purification at $D=5$ and METTS at $D=3$, and the two methods are mutually consistent within METTS' error bars.
In contrast, with lattice size increasing to $L=33,49$, the longer-range correlator $C_5$ estimated by METTS becomes noticeably stronger than its purification counterpart, and both are not quite converged in $D$. This suggests that although long-range critical correlations at finite temperature should be classical, they still require increased $D$ in METTS simulations, compare Fig.~\ref{fig:QIM_corr_L33_49_metts}.

\section{ Hubbard model}
\label{sec:Hubbard}

The Fermi-Hubbard model (FHM) is one of the simplest models of interacting fermions on a lattice, involving on-site repulsion between electrons with opposite spins. The model can be expressed as follows,
\bea
H &=& 
- \sum_{\langle i,j \rangle\sigma} 
  t\left( c_{i\sigma}^\dag c_{j\sigma} + c_{j\sigma}^\dag c_{i\sigma} \right) + \nonumber\\
  & &
  \sum_i U \left( n_{i\uparrow} - \frac12 \right) \left( n_{i\downarrow} - \frac12 \right),
\label{H}
\eea
where $c_{i\sigma}$ annihilates an electron with spin $\sigma=\uparrow,\downarrow$ at site $i$, $n_{i\sigma}=c^\dag_{i\sigma}c_{i\sigma}$ is the number operator, $n_i=n_{i\uparrow}+n_{i\downarrow}$, repulsion strength $U>0$. Here, $\langle i,j \rangle$ denotes summation over NN sites on a square lattice with hopping energy $t>0$. Despite its apparent simplicity, FHM reveals a rich array of physical phenomena, such as stripe phases and Mott insulators, due to the competition between the parameters $t$ and $U$. In one dimension, this model has exact solutions for certain limits~\cite{Lieb1968, Lieb2003}. However, obtaining thermodynamic results for a 2D system presents significant challenges, even with the most advanced numerical techniques, as discussed in the recent review by Qin~{\it et. al.}~\cite{Qin2022}.

In recent years, the study of 2D FHM at finite temperatures has seen a surge of interest. Recently, a notable body of work includes pioneering cold atom experiments~\cite{Koepsell2019, chiu19, Parsons2015, Greif2016, Cheuk2016, Parsons2016, Boll2016, Timon2017, koepsell2021microscopic, hirthe2023magnetically} where quantum gas microscopy~\cite{Bakr2009} allows high-resolution studies of many-body fermionic physics. Various finite size MPS techniques, such as exponential tensor renormalization group~\cite{Wei2021} or tangent space tensor renormalization group~\cite{li2023tangent}, have been employed to probe the rich finite temperature physics of FHM. In this context, METTS based on MPS ansatz in reduced geometries, such as thin cylinders~\cite{stripes_finite_mets, METTS_triangular}, have been particularly successful in going down to very low temperatures. Here, we present the proof-of-principle results employing genuinely 2D PEPS ansatz, focusing on a lattice at half-filling. 

\begin{figure}[t!]
\vspace{-0cm}
\includegraphics[width=\columnwidth,clip=true]{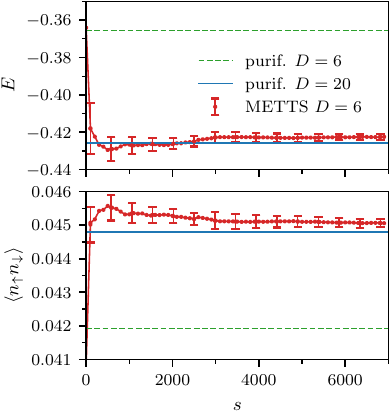}
\vspace{-0cm}
\caption{
{\bf Running average expectation values in 2D Fermi-Hubbard model.}
We show the running averages of energy $E$ in the top panel and double occupancy $\langle n_{\uparrow}n_{\downarrow} \rangle$ in the bottom panel for a $6 \times 6$ square lattice at inverse temperatures $\beta=6$, hopping $t=1$, Coulomb interaction potential $U=8$ and PEPS bond dimension $D=6$. The plots are in the function of the Markov chain's length $s$. Error bars illustrate the $99.7\%$ confidence interval. As a reference, blue lines show well-converged results from purification simulations with $D=20$, and green dashed lines represent purification outcomes for $D=6$---the same bond dimension as the one used in METTS simulations. The data reveal that METTS can approximate accurate results with significantly reduced PEPS bond dimension.
}
\label{fig:Hubbard_hf}
\end{figure}

\begin{figure}[t!]
\vspace{-0cm}
\includegraphics[width=\columnwidth]{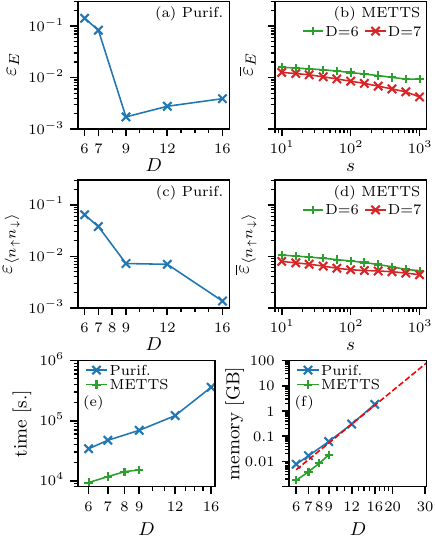}
\vspace{-0cm}
\caption{
{\bf Accuracy and numerical cost of purification and METTS.} 
The parameters of the FHM are the same as in Fig.~\ref{fig:Hubbard_hf}.
The relative errors in the first two rows are calculated with respect to well-converged purification results at bond dimension $D=20$. For purification, in panels (a) and (c), the errors are shown as a function of the PEPS bond dimension. In (b) and (d), we plot METTS error as a function of Markov chain sample size $s$. We plot mean errors here, where the average is calculated over the initial sample $s_i$, i.e., in Eq.~\eqref{eq:running_avarage}, we take here the sum over $j=s_i+1,\ldots, s_i+s$.
In (e), we show the wall-time for the imaginary time evolution of purification and a single METTS iteration. The memory bottlenecks for calculations of the expectation values are in panel (f). Extrapolating the expected $D^6$ scaling, red dashed line, employing $U(1)\times U(1)$ symmetry, it can reach about $100$GB for purification at $D=30$.
}
\label{fig:hubbard_comparisons}
\end{figure}

To expedite convergence, we employ a parallel nature of the METTS algorithm. We initialize parallel simulations starting with random product states in the occupation basis at the desired half-filling. We serialize the Markov chains data from those independent runs (four parallel runs were used in Fig.~\ref{fig:Hubbard_hf}, where maximal $s$ corresponds to total number of steps in all runs). Subsequently, we calculate the running average. To estimate statistical errors, we first bunch the data to reduce the effects of auto--correlation, and then compute a running standard deviation over the running average of the bunched data.  

Fig.~\ref{fig:Hubbard_hf} contains the running averages of the total energy per site $E$ and the expectation value of double-occupancy $n_{\uparrow} n_{\downarrow}$ for a $6 \times 6$ square lattice Hubbard model at half-filling, hopping rate $t=1$, Coulomb interaction potential $U=8$, and inverse temperature $\beta=6$.
METTS simulations with a modest bond dimension of $D=6$ closely approach converged purification results at $D=20$, showing much better convergence than the purification at $D=6$.  Simulations of FHM using iPEPS purification can be converged only for high and intermediate temperatures, limited by feasible iPEPS bond dimensions~\cite{ntuHubbard}. 
Our proof-of-principle results show that METTS with PEPS might allow overcoming those limitations, motivating further effort in this direction.

We provide further benchmarks in Fig.~\ref{fig:hubbard_comparisons}, all for the same $6 \times 6$ lattice and set of parameters.
In panels (a)-(d), we plot relative errors for energy $\varepsilon_{E}$ and double occupancy $\varepsilon_{\langle n_{\uparrow}n_{\downarrow}\rangle}$,
where $\varepsilon_{\mathcal{O}} = |(\langle \mathcal{O} \rangle - \langle \mathcal{O} \rangle_{ref} ) / \langle \mathcal{O} \rangle_{ref}|$ with the reference $\langle \mathcal{O} \rangle_{ref}$ coming from the purification simulation at $D=20$. Notably, METTS matches the accuracy of purification, but with significantly lower bond dimensions.
In Fig.~\ref{fig:hubbard_comparisons}(e), we compare a single core simulation wall-time for a single $\beta/2$ imaginary-time NTU evolution. For the same $D$, purification is taking longer here than METTS due to the presence of auxiliary degrees of freedom in PEPS tensors. METTS, however, requires multiple such iterations to achieve convergence.

Finally, Fig.~\ref{fig:hubbard_comparisons}(f) sheds light on complementary computational requirements, highlighting the rapidly increasing sizes of tensors generated during the calculation of expectation values. In particular, we show the size of the largest intermediate tensor appearing in the corner transfer matrix renormalization, zipper, or variational boundary MPS optimization algorithms. The factor of four in memory usage between purification and METTS again is related to the presence of auxlliary degrees of freedom in the former. Extrapolating the memory peak, using the expected $D^6$ scaling (red dashed line), shows that purification at $D=30$ would require memory of the order of $100$GB, putting hard limits on feasible bond dimensions. The values above are for tensors with enforced global $U(1) \times U(1)$ symmetry.

The FHM simulations have been performed using the YASTN open-source package~\cite{YASTN}. We note that to simulate fermions, we supplement the diagrams in Figs.~\ref{fig:NTU_overview}(b), \ref{fig:exp}(b), and \ref{fig:collapse}(a) with a suitable swap gates, while enforcing at least the fermionic parity symmetry, see the appendix of Ref.~\cite{ntuHubbard} for details. We also utilize a one-step version of Environment-Assisted Truncation~\cite{ntuHubbard} to enhance the initialization of the NTU optimization.

\section{Conclusion}
\label{sec:conclusion}

Simulations of thermal states by the imaginary time evolution of their purification were directly compared with a stochastic sampling and imaginary time evolution of the minimally entangled typical thermal states for systems on a finite 2D square lattice. Both the purification and METTS were represented by the same PEPS tensor network ansatz, whose imaginary-time evolution was performed with the same NTU algorithm. The divergences in implementation can solely be attributed to the essential differences between the two methods.

The comparisons were made above, at, and below the critical value of the transverse magnetic field of the 2D quantum Ising model. In all three cases, the purification method systematically required a higher PEPS bond dimension than METTS in order to reach convergence. Although METTS's lower bond dimension accelerated the imaginary time evolution simulations, this merit was overshadowed by the need for thousands of stochastic realizations. Hence, PEPS-METTS emerges as the preferable method in 2D when the bond dimension, demanded by purification, escalates to unmanageable levels, making the additional cost of sampling a justifiable trade-off. Furthermore, the sampling can be naturally parallelized, significantly expanding the range of feasible simulations. We also conducted proof-of-principle calculations for the 2D Fermi Hubbard model at half-filling, demonstrating the versatility of PEPS-METTS.

At the technical level, we introduced a zipper method to apply a row transfer matrix to the PEPS norm boundary. By deforming the boundary applying one transfer matrix tensor at a time---rather than the whole transfer matrix at once---the boundary is gradually zipped with the row matrix at a lower numerical cost.

\acknowledgements
%
A.S. thanks Alexander Wietek for discussions. This research was supported in part by the National Science Centre (TNCN), Poland under projects 2019/35/B/ST3/01028 (A.S.), 2020/38/E/ST3/00150 (M.M.R.) and project 2021/03/Y/ST2/00184 within the QuantERA II Programme that has received funding from the European Union Horizon 2020 research and innovation programme under Grant Agreement No 101017733 (J.D.).
The research was also supported by a grant from the Priority Research Area DigiWorld under the Strategic Programme Excellence Initiative at Jagiellonian University (J.D., M.M.R.).
%

\bibliography{ref.bib} 

\end{document}